\PassOptionsToPackage{table}{xcolor}
\documentclass[sigconf,nonacm]{acmart}
\AtBeginDocument{%
  }

\setcopyright{acmlicensed}
\copyrightyear{2018}
\acmYear{2018}
\acmDOI{XXXXXXX.XXXXXXX}
\acmConference[Conference acronym 'XX]{Make sure to enter the correct
  conference title from your rights confirmation email}{June 03--05,
  2018}{Woodstock, NY}
\acmISBN{978-1-4503-XXXX-X/2018/06}

\definecolor{drishticolor}{RGB}{127,0,153}   
\definecolor{denaecolor}{RGB}{30, 144, 255}  
\definecolor{martezcolor}{RGB}{255, 140, 0}  %
\definecolor{reviewcolor}{RGB}{34, 139, 34} 

\newif\ifdraft
\draftfalse        %

\ifdraft
\newcommand{\drishti}[1]{{\color{drishticolor}\textbf{[Drishti: #1]}}}
\newcommand{\denae}[1]{{\color{denaecolor}\textbf{[Denae: #1]}}}
\newcommand{\martez}[1]{{\color{martezcolor}\textbf{[Martez: #1]}}}

\newcommand{\someone}[1]{{\color{red}\textbf{[TODO: #1]}}}
\newcommand{\arxiv}[1]{{\color{red}{[\textbf{ARXIV} - #1]}}}

\else
\newcommand{\drishti}[1]{{\color{drishticolor}\textbf{}}}
\newcommand{\denae}[1]{{\color{denaecolor}\textbf{}}}
\newcommand{\martez}[1]{{\color{martezcolor}\textbf{}}}

\newcommand{\someone}[1]{{\color{red}\textbf{}}}
\newcommand{\arxiv}[1]{{\color{red}\textbf{}}}

\fi

\usepackage{verbatim}

\usepackage{xspace}
\xspaceaddexceptions{]\}}

\usepackage[table]{xcolor}
\usepackage{subfiles}
\usepackage{multirow}
\usepackage{todonotes}
\usepackage{hyperref}
\usepackage{cleveref}
\usepackage{graphicx}
\usepackage{subcaption}

\usepackage{booktabs}
\usepackage{tabularx}
\usepackage{array}
\usepackage{ragged2e}
\usepackage{xltabular}
\usepackage{multirow}
\usepackage{tcolorbox}
\tcbuselibrary{skins}
\usepackage{fontawesome5}
\newcolumntype{Y}{>{\RaggedRight\arraybackslash}X}

\newcommand{\lightmidrule}{%
    \arrayrulecolor{black!12}%
    \midrule[0.3pt]%
    \arrayrulecolor{black}%
}

\newtcolorbox{takeaway}{
  enhanced,
  colback=gray!10,
  colframe=blue!75!black,
  boxrule=0pt,
  frame hidden,
  borderline west={2pt}{0pt}{blue!75!black},
  left=5pt,
  right=4pt,
  top=3pt,
  bottom=3pt,
  sharp corners,
  before skip=5pt,
  after skip=5pt
}

\begin{document}

\title{The Labor of Storytelling: How People with Chronic Illnesses Leverage AI and Communities for Narrative Translation}

\author{Drishti Goel}
\affiliation{%
  \institution{University of Illinois Urbana-Champaign}
  \city{Urbana, IL}
  \country{USA}}
\email{drishti4@illinois.edu}
\orcid{0009-0000-6713-9240}

\author{Koustuv Saha}
\affiliation{%
  \institution{University of Illinois Urbana-Champaign}
  \city{Urbana, IL}
  \country{USA}}
\email{ksaha2@illinois.edu}
\orcid{}

\author{Martez Mott}
\affiliation{%
  \institution{Microsoft Research}
  \city{Redmond, Washington}
  \country{USA}}
\email{mamott@microsoft.com}
\orcid{0000-0003-4711-7268}

\author{Denae Ford}
\affiliation{%
  \institution{Microsoft Research}
  \city{Redmond, Washington}
  \country{USA}}
\email{denae@microsoft.com}
\orcid{0000-0003-0654-4335}

\renewcommand{\shortauthors}{Goel et al.}

\begin{abstract}

The distinctiveness of people with chronic illness as storytellers is how they experience their illness. Symptoms, pain, and biological disruptions are eloquent forms of speech of the body that beget social articulation. The stories people tell about illness shape how suffering is legitimized, and acted upon. Large language models have emerged as new mediators within this landscape---a role once predominantly supported by communities. In doing so, these systems participate in what we term \textit{narrative translation}: transforming lived illness experiences for particular audiences and purposes. Through 21 prototype-driven semi-structured interviews with people with chronic illness, we contribute an empirical account of how participants navigate this mediation alongside existing sources of support. Our findings surface distinct yet complementary roles participants envisioned for AI and their communities, and how trust in AI was shaped by learning model-boundaries and perceiving understanding. We characterize these findings to inform future designs of AI-mediated narrative support.

\end{abstract}


\maketitle

\section{Introduction} 

\begin{flushleft}
\emph{``The stories we tell about our lives are not necessarily those lives as they were lived, but these stories become our experience of those lives.''}
--Arthur W. Frank~\cite{frank2013wounded}
\end{flushleft}

\noindent Chronic illness changes the temporal logic of care: the work of explaining, adapting, and being understood~\cite{aarhus2010negotiating, corbin1991nursing, corbin1998corbin, gbd2025characterising}. Rather than a bounded episode that ends with recovery, chronic conditions persist over time, often with no cure, requiring ongoing treatment, symptom management, self-management, and adaptation to changing capabilities and circumstances~\cite{cdcChronicDisease2026, thorne2000two}. In 2023, 76.4\% of U.S. adults reported living with at least one of twelve measured chronic conditions, while 51.4\% reported multiple conditions~\cite{watson2025trends, national1999health}. Scholarship has long characterized the consequences of this persistence through concepts such as illness trajectories, biographical disruption, loss of self, and the ``unending work'' of living with illness~\cite{corbin1988unending,bury1982chronic,charmaz1983loss, strauss1976chronic, charmaz1991good}. Chronic illness, therefore, involves not only living with an enduring condition of the body, but continually accommodating its effects within everyday routines, relationships, identities, and expectations for the future.
The embodied experience of illness is often not directly accessible to others, and medical sociology and anthropology research has often distinguished between the lived and embodied experience of illness and how people describe and communicate that experience to others~\cite{kleinman2020illness, good1994medicine, palla2024narrative,charmaz2002stories,williams1984genesis}. Whether a person's account of their illness is believed, in turn, has direct consequences beyond the clinical encounter itself: it shapes whether they receive necessary workplace accommodations, disability support, institutional recognition of their condition, which in turn defines their societal experience of the illness~\cite{frank2013wounded, gignac2021disclosure, loseke2011empirical, goffman1978presentation}. 
Communication is not merely ancillary to the management of chronic illnesses, but also one of the means through which people make their illness socially and institutionally actionable.

Historically, this narrative labor has largely fallen on the person with chronic illness, drawing on their prior experience, trial and error, or resources such as clinicians or patient navigators, peer communities and online health forums, where people exchange not just medical information but strategies for articulating symptoms~\cite{mamykina2015collective, zhao2022online, jacobs2019think, berry2019supporting}. This self-directed information- and strategy-seeking reflects a broader shift toward patients developing ``lay expertise'' about their own conditions~\cite{sosnowy2014practicing}. 
More recently, generative AI and large language models (LLMs) have emerged within this landscape, further compounding these preferences. These people are increasingly turning to AI tools to draft messages to clinicians, prepare for appointments, process medical information, and rehearse self-advocacy conversations~\cite{yang2024wish, yang2024talk2care}. Emerging HCI and health-informatics literature has begun to document both the promise of this shift; by studying AI's capacity to bridge knowledge and literacy gaps between patients and providers~\cite{yang2024wish, yun2025online}, and the looming tensions around trust, reliability, and the erosion of existing human support networks~\cite{yang2024wish, ma2024understanding, draxler2024ai, gero2023social, dhillon2024shaping, joshi2025writing}. 
In this study, we seek to understand how people with chronic illness employ AI in what we term \emph{narrative translation}: the recurring work of transforming lived illness experience into forms that are intelligible and appropriate to different audiences---such as clinicians, employers, family members, or online communities---each with different expectations around evidence, tone, and disclosure. We use \textit{translation} deliberately. Building on the distinction between the experiential ``voice of the lifeworld'' and the institutional ``voice of medicine''~\cite{mishler1984discourse}, illness communication can be understood as movement between distinct interpretive regimes. Rendering embodied experience as a clinical symptom, for instance, does not merely paraphrase it; it foregrounds particular meanings and makes them legible and actionable within biomedical practice~\cite{wardrope2022hermeneutics}. We extend this theoretical lens beyond the clinical setting as people with chronic illness repeatedly translate experience across social and institutional contexts.
Our study was guided by three research questions: 

\begin{enumerate}
    \item[\textbf{RQ1:}] How do people with chronic illness use AI to mediate the process of narrative translation?

    \item[\textbf{RQ2:}] How do people with chronic illness negotiate trust and reliance when AI participates in interpreting and translating their experiences?

    \item[\textbf{RQ3:}] How, if at all, do these people leverage support from communities for communicating these illness experiences?

\end{enumerate}

To investigate these dynamics, we conducted a technology probe-driven semi-structured interview study with 21 people living with chronic illness. Our findings show that narrative translation begins before an account is written and continues beyond its reformulation for a particular audience. Participants used AI to organize and recall experiences, anticipate how
different audiences might respond, and rehearse consequential interactions. They moved between AI and lived-experience communities for different forms of support, while continually evaluating what information and interpretations they were willing to accept, revise, or carry forward. Across these practices, AI did not simply assist in expressing an illness narrative; it became involved in determining how experience was interpreted, contextualized, and represented. Our contributions are summarized as follows:

\begin{enumerate}

    \item We conceptualize \textit{narrative translation} as the recurring, situated work through which people with chronic illness render lived experience communicable across audience, and empirically characterize how AI participates in this work through sensemaking, audience adaptation, anticipation, and calibrated reliance.

    \item We characterize how participants negotiated \textit{trust and reliance} through repeated interactions with AI, showing how accumulated trust did not uniformly translate into reliance and how reliance varied across tasks and participants, from primarily instrumental to more relational orientations.

    \item We characterize AI and lived-experience communities as distinct, but, potentially complementary resources for illness sensemaking and communication, and identify a bidirectional design space for integrating computational and experiential forms of support. 

\end{enumerate}

Together, these contributions foreground the need toward understanding how AI participates in the ongoing work through which illness experiences are interpreted, translated, and made consequential across contexts. In doing so, we highlight opportunities for AI systems that support movement across audiences and sources of lived experience while preserving people's authority over how their experiences are understood and represented.

\section{Related Work}

\subsection{Illness Narratives as Situated Translation Work}

A long line of work separates \textit{disease}, the biomedical condition, from \textit{illness}, the meanings people attach to symptoms and suffering~\cite{kleinman1978culture,kleinman2020illness}. Bury and Charmaz extend this to argue that chronic illness disrupts biography and self-concept, not just the body~\cite{bury1982chronic,charmaz1983loss}. Building on this, prior work also treats storytelling as the mechanism by which people render disruption intelligible: illness narratives organize experience into forms others can interpret, not just report events~\cite{frank2013wounded}. Charon extends this into clinical practice as \textit{narrative competence}: clinicians' capacity to receive patients' stories as stories, rather than translating them immediately into symptoms~\cite{charon2001narrative}. 

Another line of work extends this by showing that the listener's frame is not neutral. Mishler's ``voice of the lifeworld'' versus ``voice of medicine'' distinction shows that a patient's contextual accounts are routinely reinterpreted through institutional, medically legible categories~\cite{mishler1984discourse, barry2001giving}. Thorne et al. similarly find that communication needs vary by ``disease world''---whether a condition is visible, socially legitimate, or treatable---meaning no single communication strategy transfers across conditions~\cite{thorne2004context}. Consistent with this, Iroegbu et al.'s systematic review of 13 studies identifies \textit{adaptive interpersonal communication}, tailoring communication to a patient's specific circumstances, as the throughline of what makes patient-provider communication effective for self-management~\cite{iroegbu2025influence}. Hayden goes further, showing that misalignment is not limited to clinical encounters: her interviews with chronically ill participants found that they felt structurally misunderstood by healthy people, since illness altered the meaning of even ordinary words~\cite{hayden1993chronically}. Ganesh and Lazar framed the resulting behavior as ongoing \textit{disclosure work}---deciding whether, when, and how much of a condition to make legible to a given audience~\cite{ganesh2021work}.

These accounts establish that translation across audiences is effortful and consequential, but leave open questions about \textit{how} people actually perform this translation over time, or what is negotiated when a third party participates in producing the narrative. We refer to this process as \textit{narrative translation}: transforming lived illness experience into forms legible to a given audience.

\subsection{Community \& AI-Mediated Health Communication and Narrative Support}

For people with chronic illness, learning how to communicate an experience has historically been an interpersonal endeavor, supported by peers~\cite{hartzler2011managing, civan2009locating}, in-person support groups~\cite{borkman1976experiential, davison2000talks}, clinicians~\cite{werner2003hard, iroegbu2025influence}, and patient navigators~\cite{freeman2011history, jacobs2014cancer}. Peers, in-person support groups, clinicians, and patient navigators were joined, from the late 1980s onward, by computer-mediated self-help groups hosted on bulletin boards, Usenet, and mailing lists~\cite{finn1994computer, gustafson1993development, loader2002medicine}. The subsequent growth of the internet and social platforms expanded these into large, searchable, condition-specific communities~\cite{ziebland2012health, zhao2022online, gatos2021hci, kaliappan2025online, levonian2021patterns}. Increased connectivity,
varying degrees of anonymity, and access to others with the same condition, contributed toward people with chronic illness seeking informational, emotional, and relational support in these spaces~\cite{sosnowy2014practicing, naslund2016future, johnson2022s, jacobs2019think}.

Borkman characterizes what circulates in such settings as \textit{experiential knowledge}: authority grounded in having undergone a condition rather than in professional training~\cite{borkman1976experiential}, a distinction Hartzler and Pratt make empirically in showing that patient expertise addresses dimensions of living with illness that clinical expertise does
not~\cite{hartzler2011managing, civan2009locating}. Ziebland and Wyke identify \textit{learning to tell the story} as one of the mechanisms through which sharing experiences online affects health, placing narrative rehearsal alongside information seeking and social support~\cite{ziebland2012health}. 

While earlier mediators carried, hosted, or received an account; a generative system can participate in composing it. Hancock et al. define \textit{AI-mediated communication} as communication in which an intelligent agent modifies, augments, or generates a message on a communicator's behalf~\cite{hancock2020ai}. The responses generated by humans and AI can also differ significantly in terms of their linguistic traits and perceptions~\cite{saha2025ai, hohenstein2023artificial, jakesch2023co}. 
Jakesch et al., in an online experiment with 1,506 participants, found that opinionated language-model suggestions shift not only what people write but the views they subsequently report~\cite{jakesch2023co}. 
Moreover, in healthcare, this mediation has largely been studied as a documentation and decision-support aid. Kim et al.'s \textit{MindfulDiary}, evaluated in a four-week field study with 28 patients with major depressive disorder and five psychiatrists, how LLM prompting helped patients document daily experiences while surfacing those records to clinicians through a dashboard~\cite{kim2024mindfuldiary}. Among a long line of work, Cao et al.'s diary study of 25 patients similarly frames general-purpose and LLM use as spanning informational, behavioral, and emotional needs, proposing LLMs as ``boundary companions'' between patients and clinicians~\cite{cao2026more, lucchini2026self, chen2023soulchat, sadilek2012modeling, zheng2023building}. 
Together, these studies establish that AI can participate in producing an account, but leave open the question of how people with chronic illness use it to translate their experiences for audiences beyond the clinic, how they decide to trust it with those accounts, and what they continue to seek from communities instead.

\section{Methods}
\subsection{Study Context}

We study \textit{narrative translation} as the ongoing, situated work through which people render lived experiences intelligible to others across different relational and institutional contexts. For people living with chronic illness (PwCI), this work can involve communicating experiences that are embodied, subjective, and not always readily observable or legible to others. Prior work in medical sociology and anthropology has similarly emphasized that illness is not reducible to biomedical disease, but is experienced through disruptions to everyday life, identity, relationships, expectations, and systems of meaning \cite{bury1982chronic,kleinman2020illness}. We build on this orientation to conceptualize narrative translation not as the linguistic reformulation of a stable underlying story, but as a continuous process of deciding \textit{what} about an experience should be communicated, \textit{how} it should be framed, and \textit{for whom} it is being made intelligible.

Narrative translation is therefore inherently situated. A person's account may change when communicating with a clinician, employer, family member, friend, or peer community because each audience brings different relationships, expectations, forms of expertise, and potential consequences. Translation can consequently involve decisions about what to disclose or withhold, what aspects of an experience to foreground, what vocabulary or tone to adopt, and how much an account can be transformed while continuing to represent the narrator's experience. We use ``translation'' deliberately to capture this movement between experiential and communicative contexts, rather than to imply a one-to-one conversion between two fixed versions of the same story.

To investigate this process empirically, we grounded each interview in a personally relevant communication situation identified by the participant before the session. Participants described an experience involving their chronic illness and a particular person or audience with whom they needed, or had needed, to communicate. We returned to this same situation throughout the interview, allowing participants to reason about AI assistance in relation to a stable set of communicative stakes---including the intended audience, their relationship to that audience, what they hoped the recipient would understand, and what they considered consequential about being understood or misunderstood. This approach enabled us to examine participants' judgments in relation to experiences they themselves considered meaningful rather than through decontextualized or researcher-authored scenarios.

We were particularly interested in how generative AI enters this situated work alongside other resources participants already draw on, including their own prior experience, people in their personal networks, and lived-experience communities. Rather than presupposing AI to be the primary or preferred source of support, our interviews examined how participants combined, distinguished, or moved between these resources and what forms of support they attributed to each. This allowed us to consider AI-assisted narrative translation within a broader ecology of interpretive and relational support rather than as an isolated human--AI interaction.

To make otherwise tacit aspects of this process available for reflection, we additionally introduced a theory-informed technology probe during the interviews. Following the tradition of technology probes as deliberately exploratory artifacts used to learn about practices, needs, and possible technologies rather than to evaluate a finished system \cite{hutchinson2003technology}, the probe instantiated several possible forms of AI mediation for participants to encounter and critique. Its purpose was not to solicit feature-level feedback for the probe itself. Instead, its intentionally partial and contestable design created concrete occasions for participants to articulate what existing AI systems fail to support, what forms of mediation they would accept or resist, and how they imagined AI might differently participate in narrative translation.

\subsection{Study Design}

\subsubsection{Screening and Pre-Interview Survey}
Interested participants completed an online screening and pre-interview survey administered through Qualtrics (see ~\autoref{sec:Appendix_survey}). The survey assessed eligibility (age, U.S. residence, and chronic-illness experience) and collected demographic information, including age, gender, race/ethnicity, and employment status. To contextualize participant's experiences, we additionally asked about their chronic condition(s), duration and perceived visibility of illness, frequency and contexts of illness-related communication, sources they typically turned to for support, and their prior use---or non-use---of AI for illness-related communication. Participants were also asked, in advance of the interview, to identify and
briefly describe a personally relevant situation in which they had to communicate about their health but experienced uncertainty about how to frame it, including the intended audience, communicative goal, source of difficulty, and whether they had used AI or other support. The interviewer reviewed this response before each session, and the participant-selected situation was used across subsequent interview activities to ground discussion and interaction with the technology probe in a concrete audience, relationship, and communicative goal. Eligible participants reviewed and provided consent through the survey before scheduling a 90-minute interview. The average time taken to complete this form was about 24 minutes.

\subsubsection{Probe-Driven Interview}
We conducted approximately 90-minute remote semi-structured interviews consisting of three phases: a naturalistic walkthrough, interaction with a technology probe, and reflective envisioning. We first returned to the communication situation participants had described in the pre-interview survey and asked them to elaborate on the experience, including what they wanted their intended audience to understand, what made the communication difficult, and how they navigated these challenges. Participants then demonstrated or reconstructed how they used---or would use---AI and other sources of support in this process, allowing us to ground the interview in their existing practices before introducing our own design. Next, participants worked through the same communication situation using a theory-informed technology probe while thinking aloud about what they found useful, inaccurate, uncomfortable, unexpected, or missing. Drawing on technology probes as exploratory artifacts for understanding users' needs and inspiring possibilities for future technologies~\cite{hutchinson2003technology}, we used the probe to materialize alternative forms of AI-mediated narrative translation for participants to inspect, challenge, and reimagine, rather than to evaluate it as a finished system. Participants were asked to share their screen as they interacted with the probe. Finally, we moved into reflective envisioning, where participants compared their experiences across existing tools and the probe and reflected more broadly on sensemaking, authorship and agency, trust, personalization, community support, and boundaries for future AI systems. Throughout the interview, we followed a semi-structured protocol while allowing participant's experiences and reactions to guide follow-up questions and discussion. \textit{For example, we asked participants to compare how it felt for an AI system to state that others experienced similar symptoms with hearing the same sentiment from someone sharing lived experience in an online support community; to articulate the limits they believed AI should observe when expressing empathy or relational understanding; and to reflect on experiences that had led them to avoid, discontinue, or increasingly rely on particular AI tools.} Prior to the main study, we piloted the procedure with three participants to refine both the interview protocol and the use of the technology probe as an elicitation tool (see~\autoref{sec:Appendix_pilots})

\subsection{Technology Probe Construction}
\label{subsec:probe}

We developed a theory-informed technology probe, to make possible forms of AI-mediated narrative translation concrete enough for participants to experience, critique, and reimagine. Technology probes have been used in HCI to elicit needs and practices while provoking reflection about possible future technologies~\cite{hutchinson2003technology}. Accordingly, the probe was not designed as a finished system for usability or feature evaluation. Instead, it served as an elicitation artifact through which participants could relate particular forms of AI mediation to their existing practices and articulate what they would retain, resist, or envision differently.

It combined theoretical perspectives motivating our study with design mechanisms around transparency, user control, and access to lived experience identified as important in prior HCI research~\cite{abdul2018trends, amershi2019guidelines}. We iteratively developed and refined these mechanisms through discussions among the research team, considering how each could instantiate the study's theoretical framing without prescribing a single model of AI-supported communication. The resulting probe comprised four main stages: \textit{intake and interpretation}, \textit{audience and style configuration}, \textit{narrative translation and review}, and \textit{community experiences}.

\textbf{Intake and interpretation.}
In this stage of the probe, participants described their illness experiences and the context of the intended conversation to a conversational chatbot. The intake drew on seven dimensions adapted from Kleinman's Explanatory Model~\cite{kleinman2020illness}. The system then surfaced model-extracted statements from the conversation for participants to review before choosing whether to add them to the model's \textit{Knowledge Base}.

\textbf{Audience and style configuration.}
After providing context to the model and reviewing the \textit{Knowledge Base}, participants selected their intended audience from a set of predefined scenarios (e.g., workplace, clinical, or personal contexts) or specified a custom audience. They then adjusted the \textit{Style Console} to configure how their narrative would be communicated, adjusting characteristics ranging from tone and level of detail to the framing of agency in the final response.

\textbf{Narrative translation and review.}
Finally, the probe generated three alternative narratives informed by Frank's Restitution, Chaos, and Quest Narrative Forms~\cite{frank2013wounded}. We used these forms as design resources for producing different representations, rather than as classifications of participants or their experiences. Alongside these alternatives, the interface presented model-generated assessments of audience fit and fidelity to the participant's account; a \textit{Didn't Fit} panel showing information omitted from each narrative and the model's rationale. Together, these mechanisms exposed otherwise implicit decisions about interpretation, transformation, and omission for participants to examine and contest. 

\textbf{Community experiences.}
The interface also surfaced four simulated community posts related to the participant's account. These posts were model-generated rather than retrieved from actual community platforms, allowing us to instantiate AI--community interactions without using real community data while maintaining control over potentially sensitive content surfaced to participants. Participants used these posts to reflect on how lived-experience narratives might complement, differ from, or be integrated with AI-mediated support.

Across these stages, the probe instantiated different points at which AI might mediate narrative translation: scaffolding the articulation of context, surfacing and enabling review of the model's knowledge about a participant's experience, configuring audiences and communication styles, generating alternative narrative representations, transparently omitting information, and connecting participants with relevant lived experiences from communities. We intentionally made these processes visible and revisable to elicit participants' reflections on where and how AI should participate in this work. Therefore, we did not collect numerical ratings of individual probe-features or treat participants' responses as evaluations of system performance. Instead, participants were encouraged to critique the probe, relate its forms of mediation to their existing practices, and envision alternative forms of AI-mediated support.

\begin{table*}[t]
    \centering
    \caption{Probe-features, motivating ideas, and literature informing their construction.}
    \label{tab:probe-grounding}

    \small
    \setlength{\tabcolsep}{6pt}

    \begin{tabularx}{\textwidth}{
        >{\raggedright\arraybackslash}p{0.25\textwidth}
        >{\raggedright\arraybackslash}X
        >{\raggedright\arraybackslash}p{0.17\textwidth}
    }
        \textbf{Probe Feature} &
        \textbf{Motivating Ideas / Probe Points} &
        \textbf{Relevant Literature} \\
        \midrule

        Conversational Intake &
        How might AI elicit embodied, personal, and social dimensions of illness? &
        ~\cite{kleinman2020illness, charon2001narrative, wardrope2022hermeneutics} 
        \\[0.35em]
        \lightmidrule

        Knowledge Base Review &
        What has AI interpreted or inferred from a person's account, and how might these interpretations be used? &
        ~\cite{lim2009and, dey2009support, amershi2019guidelines, alfrink2023contestable}
        \\[0.35em]
        \lightmidrule

        Audience \& Style Console &
        How would people, if at all, want to control the tone and nature of their experience as presented by AI? &
        ~\cite{bell1984language, mishler1984discourse, hancock2020ai, iroegbu2025influence}
        \\[0.35em]
        \lightmidrule

        Alternative Narrative Translations &
        Which factors contribute to the evaluation and preference for AI responses? &
        ~\cite{frank2013wounded, charon2001narrative, wardrope2022hermeneutics}
        \\[0.35em]
        \lightmidrule

        Audience Fit, Fidelity \& \textit{Didn't Fit} &
        What is transformed, foregrounded, or omitted through AI-mediated translation, and how might these decisions remain available for examination
        and revision? &
        ~\cite{hancock2020ai, lim2009and, alfrink2023contestable, jakesch2023co}
        \\[0.35em]
        \lightmidrule

        AI-Generated Community Accounts &
        How might others' experiences support reflection and sensemaking, and how does experiential knowledge differ from other sources of support? &
        ~\cite{civan2009locating, hartzler2011managing, huh2014weaving, ziebland2012health, gatos2021hci}
        \\

        \bottomrule
    \end{tabularx}
\end{table*}

\begin{figure*}
    \centering

    \begin{subfigure}[t]{0.30\textwidth}
        \centering
        \includegraphics[
            width=\linewidth,
            height=0.31\textheight,
            keepaspectratio
        ]{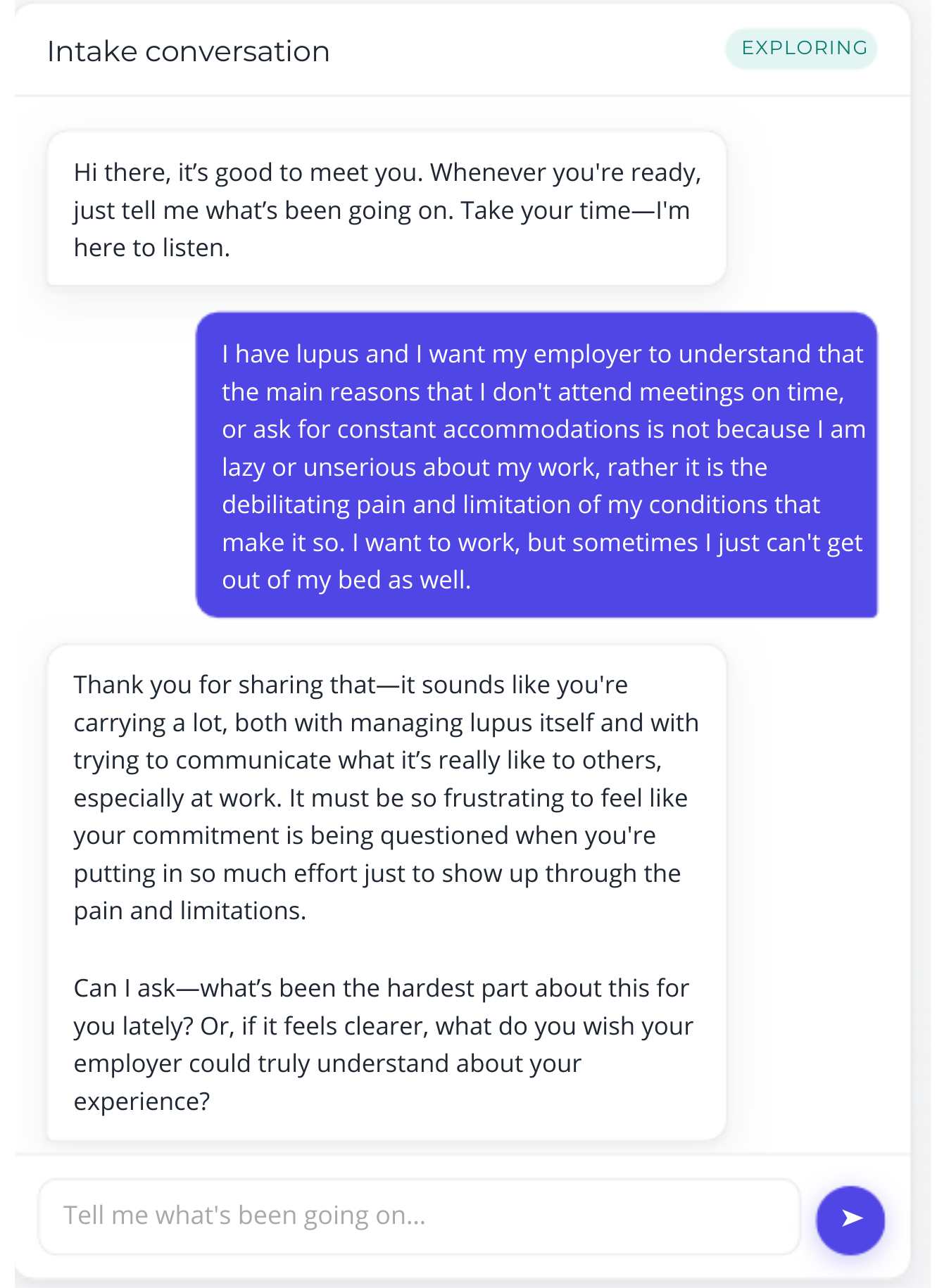}
        \caption{\textbf{Conversational Intake}}
        \label{fig:probe-articulate}
    \end{subfigure}
    \hfill
    \begin{subfigure}[t]{0.30\textwidth}
        \centering
        \includegraphics[
            width=\linewidth,
            height=0.31\textheight,
            keepaspectratio
        ]{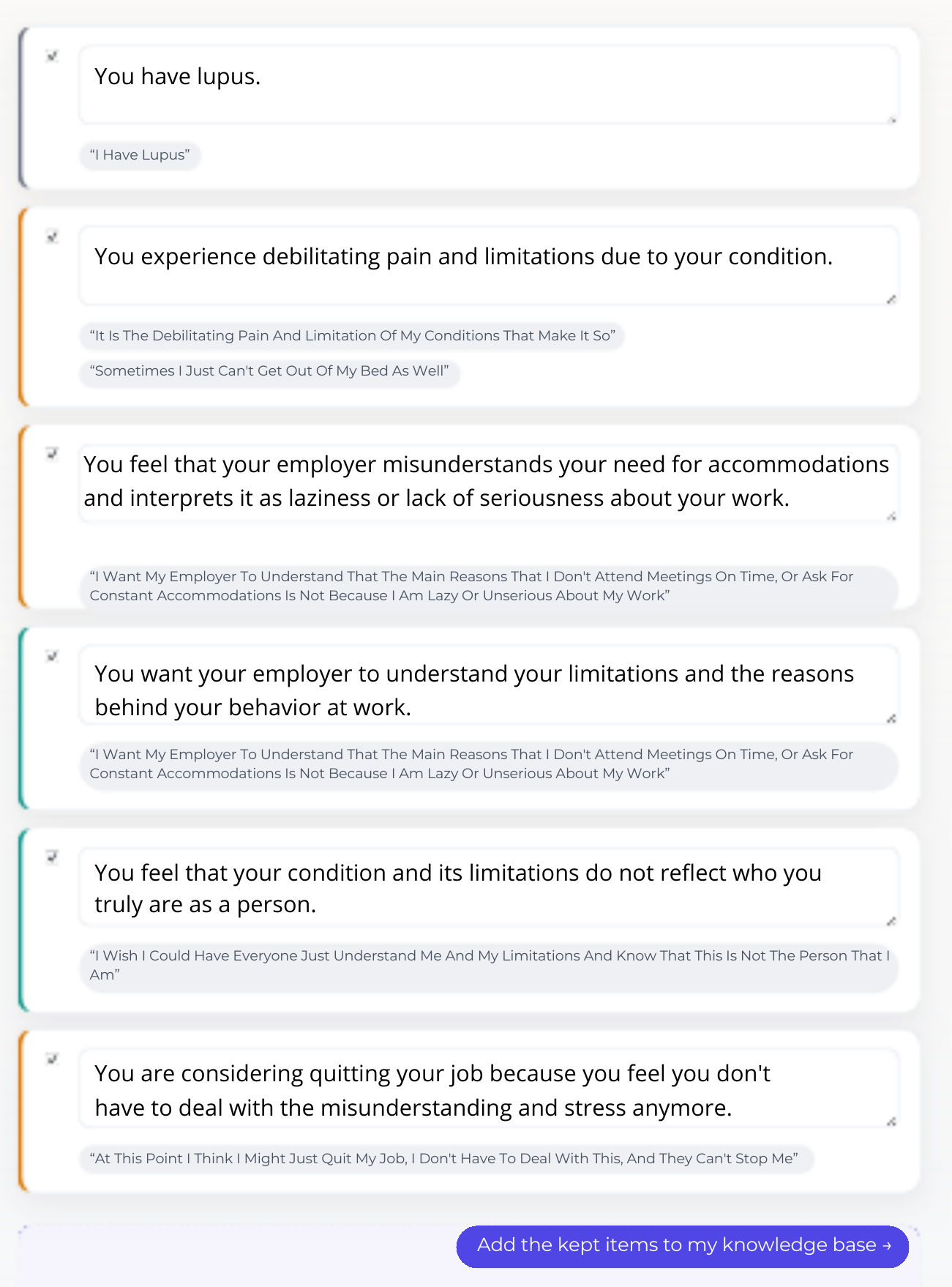}
        \caption{\textbf{Knowledge Base Review}}
        \label{fig:probe-verify}
    \end{subfigure}
    \hfill
    \begin{subfigure}[t]{0.30\textwidth}
        \centering
        \includegraphics[
            width=\linewidth,
            height=0.31\textheight,
            keepaspectratio
        ]{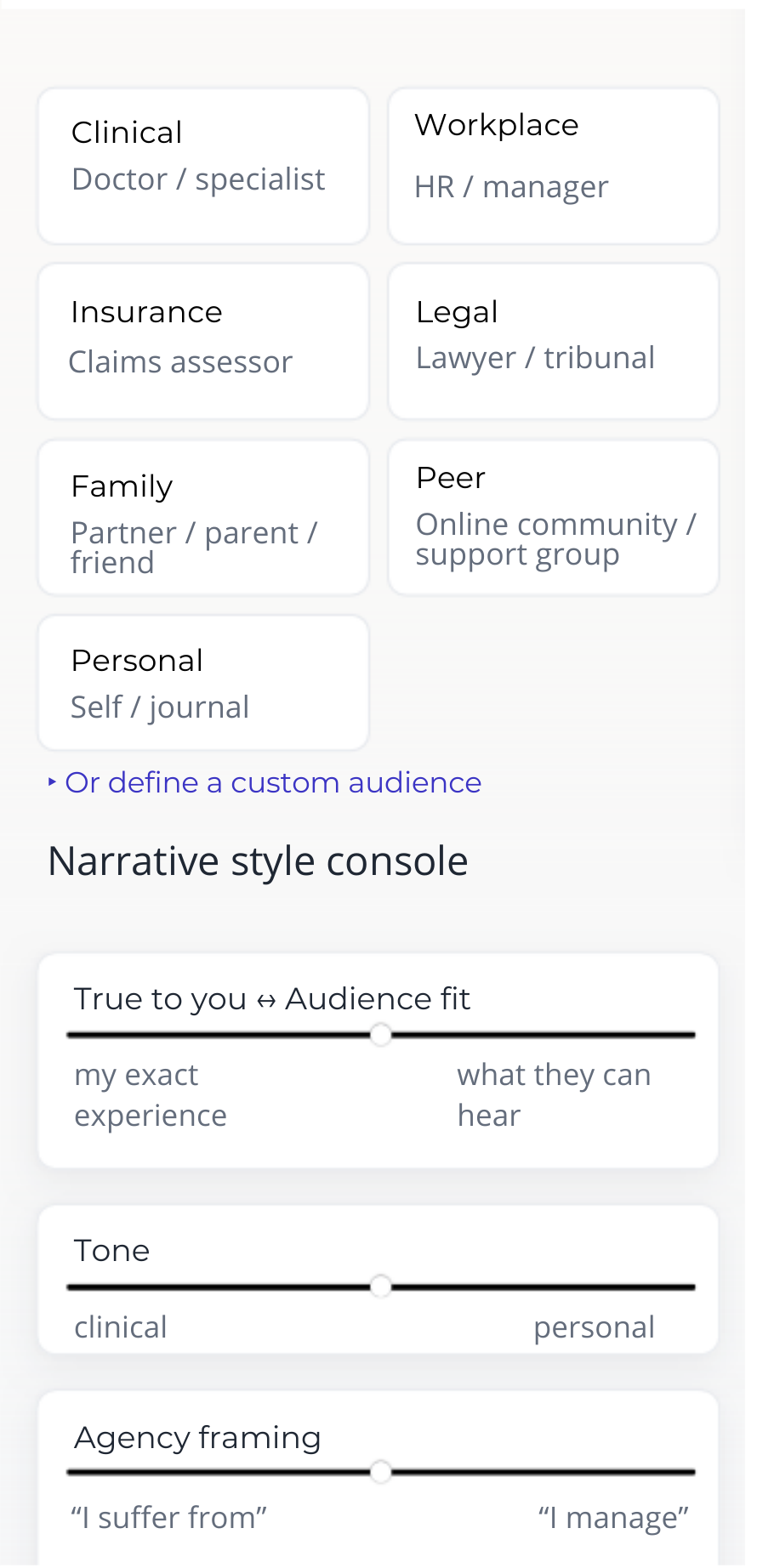}
        \caption{\textbf{Audience Style}}
        \label{fig:probe-configure}
    \end{subfigure}

    \caption{
        \textbf{Constructing and configuring a narrative for translation using the technology probe.}
        Participants first articulated an illness experience, reviewed and
        verified the probe's representation of that experience, and configured
        the intended audience and narrative framing before translation.
    }
    \label{fig:probe-construction}
\end{figure*}

\begin{figure*}
    \centering

    \begin{subfigure}[t]{0.49\textwidth}
        \vspace{0pt}
        \centering
        \includegraphics[
            height=0.43\textheight
        ]{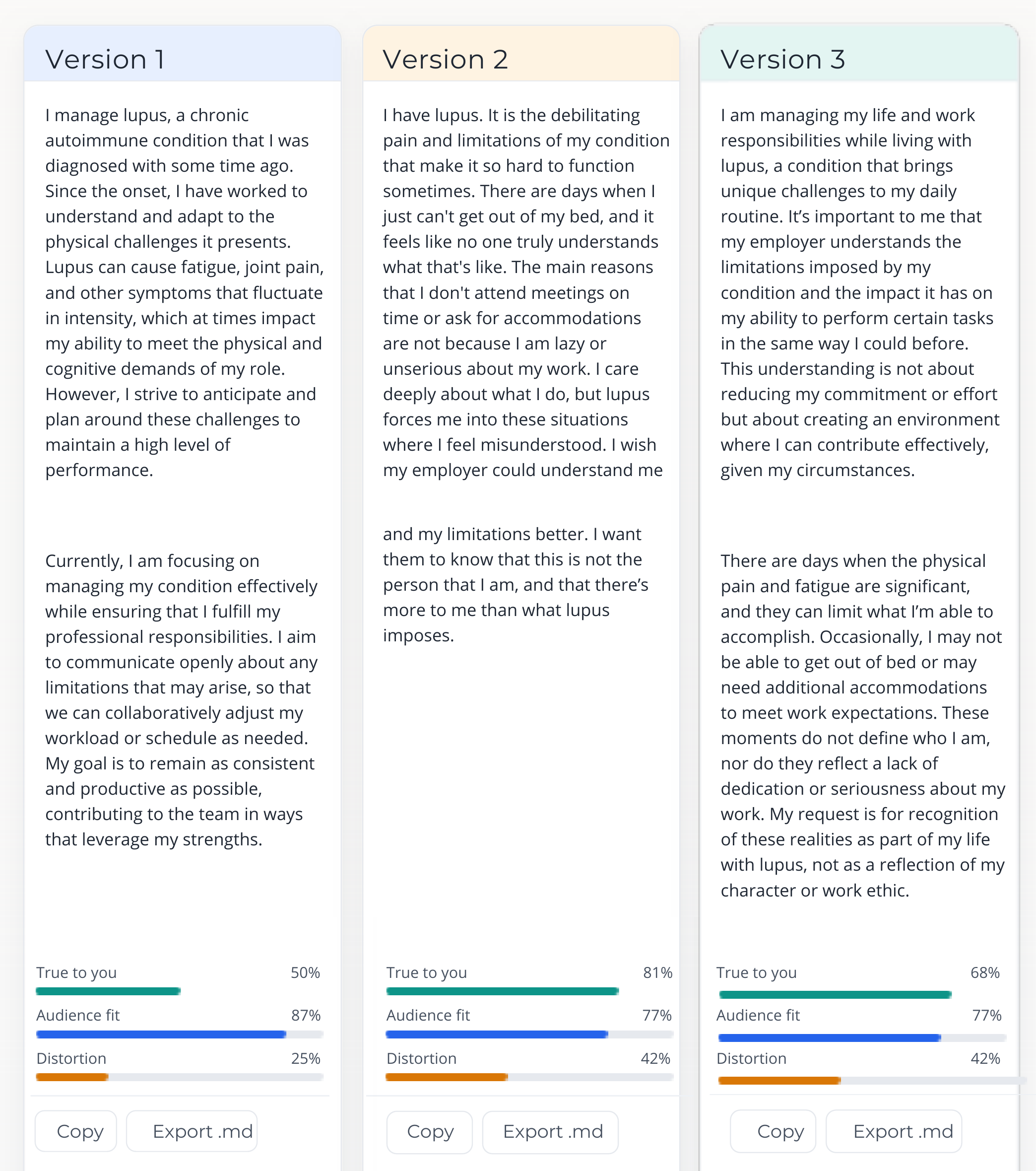}
        \caption{\textbf{Translate}}
        \label{fig:probe-translate}
    \end{subfigure}
    \hfill
    \begin{subfigure}[t]{0.49\textwidth}
        \vspace{0pt}
        \centering
        \includegraphics[
            height=0.43\textheight
        ]{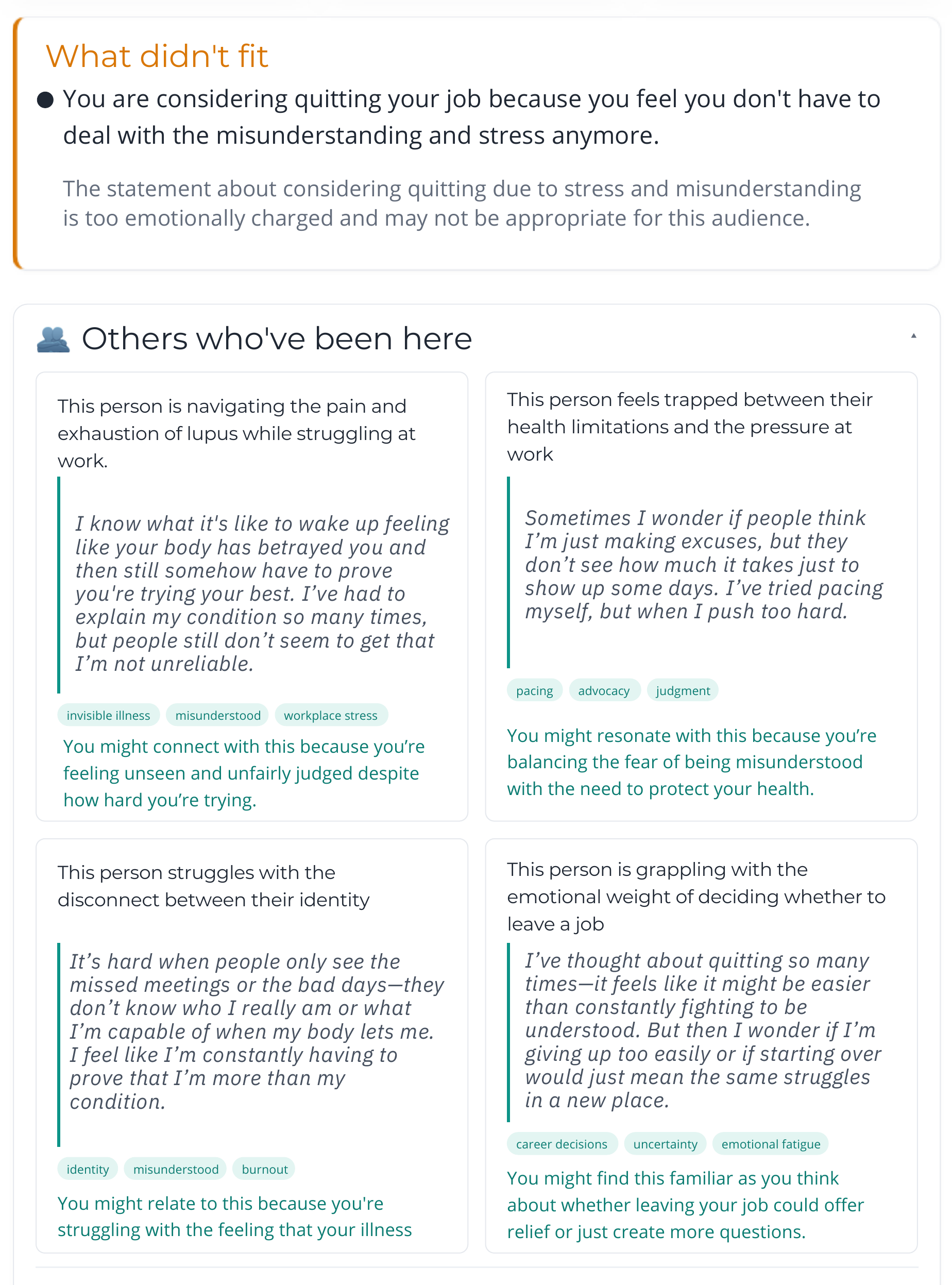}
        \caption{\textbf{Reflect and connect}}
        \label{fig:probe-community}
    \end{subfigure}

    \caption{
        \textbf{Translation and reflection beyond the translated narrative in the technology probe}
        Participants compared alternative audience-specific translations and
        their trade-offs in fidelity, audience fit, and distortion. The probe
        also surfaced content that did not fit the selected translation and
        composite accounts of related lived experiences, enabling participants
        to reflect on narrative loss and resonance with others.
    }
    \label{fig:probe-translation-reflection}
\end{figure*}

\subsection{Recruitment and Participants}

We recruited participants through Prolific over a three-week period. Eligible participants were adults 18 years or older residing in the United States who reported lived experience with chronic illness, were comfortable participating in an interview conducted in English, and had access to a computer and stable Internet connection to support video conferencing and screen sharing. Recruitment materials directed interested participants to the screening and pre-interview survey described above. Participants who met the eligibility criteria, provided informed consent, and completed the survey were invited to schedule a 90-minute remote interview. Recognizing that chronic illness can involve fluctuating symptoms, fatigue, and other constraints that may affect participation, we allowed participants to reschedule interviews as needed and explicitly encouraged them to take breaks during the session. Participants received \$115 upon completion of the study. The study protocol was reviewed and approved by our organization's Institutional Review Board.

\begin{figure}
\centering
\includegraphics[width=\linewidth]{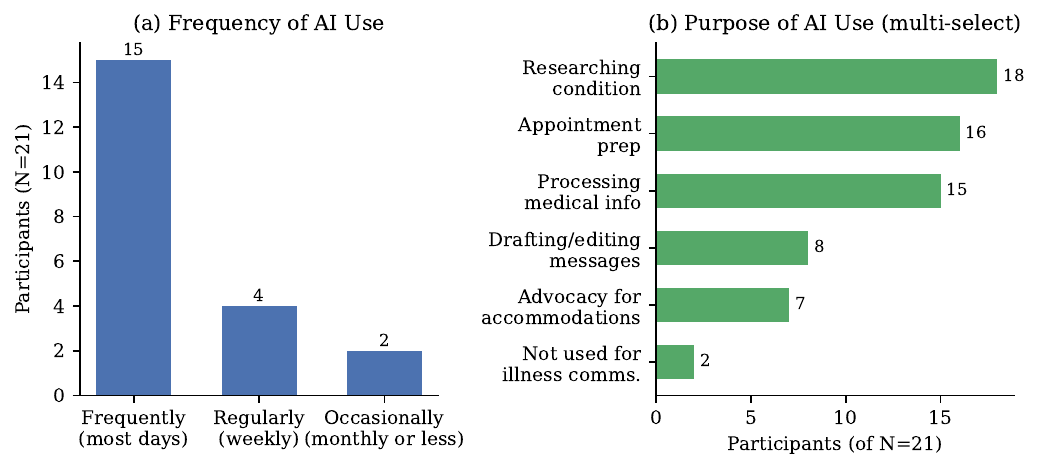}
\caption{Frequency and purpose of AI use for illness-related communication among participants (N=21). Panel (b) reflects multi-select responses, so counts exceed N.}
\Description{Two-panel bar chart. Panel (a) shows frequency of AI use among 21 participants: 15 use AI frequently (most days), 4 regularly (weekly), and 2 occasionally (monthly or less). Panel (b) shows purpose of AI use, a multi-select item: researching condition (18), appointment prep (16), processing medical info (15), drafting or editing messages (8), advocacy for accommodations (7), and not used for illness communication (2).}
\label{fig:ai_use}
\end{figure}

Our final sample consisted of 21 participants, ranging in age from 18 to 64 years, self-identified as women (n=12, 57\%), men (n=7, 33\%) and non-binary (n=2, 10\%). Participants reported living with a range of chronic conditions, and many reported multiple co-occurring diagnoses. Table~\ref{tab:demographics} summarizes participant demographics and illness details. Participants also varied in their prior use of AI for illness-related communication (Figure~\ref{fig:ai_use}). 
We identify participants using pseudonymous IDs (P1--P21) throughout the paper.

\begin{table}[htbp]
\centering
\footnotesize
\caption{Summary of participants, including Reported Illness, Gender, Age, and Years with Illness (YwI).}
\label{tab:demographics}
\begin{tabular}{@{}lllll@{}}
\toprule
ID & Illness & Gender & Age (yrs.) & YwI (yrs.)  \\
\midrule
P1  & Gastroparesis, Achalasia, GERD        & Woman     & 35-44 & 10+   \\
P2  & Lupus                                  & Woman     & 18-24 & 1--2\\
P3  & Celiac, Hashimoto's, AS                & Woman     & 24-34 & 6--10 \\
P4  & GERD / GI issues                       & Woman     & 45-54 & 6--10 \\
P5  & IBS-D / Crohn's-suspected               & Woman     & 45-54 & 3--5 \\
P6  & Crohn's, IBS, cancer survivor          & Woman     & 35-44 & 3--5 \\
P7  & Sj\"{o}gren's, Hashimoto's, FMS, Lupus & Woman     & 45-54 & 6--10 \\
P8 & Myasthenia Gravis, Lupus               & Woman     & 35-44 & 6--10 \\
P9 & IBS / SIBO                             & Man       & 35-44 & 3--5 \\
P10 & Beh\c{c}et's, Hidradenitis Suppurativa & Man       & 24-34 & 10+ \\
P11 & PCOS, CFS, Endometriosis, Asthma       & Woman     & 45-54 & 10+ \\
P12 & Ehlers-Danlos, Crohn's                 & Woman     & 24-34 & 10+ \\
P13 & Fibromyalgia, Lupus                    & Woman     & 35-44 & 6--10 \\
P14 & Endometriosis                          & Woman     & 45-54 & 10+ \\
P15 & Chronic Diarrhea (poss. Crohn's/CF)    & Man       & 45-54 & 10+ \\
P16 & Epilepsy, autoimmune, TBIs             & Non-binary & 24-34 & 10+ \\
P17 & Chronic Pancreatitis                   & Man       & 35-44 & 10+ \\
P18 & ADHD, OCD, Tourette's                  & Man       & 35-44 & 10+ \\
P19 & Fibromyalgia / CFS                     & Non-binary & 24-34 & 10+ \\
P20 & Rheumatoid Arthritis                   & Man       & 55-64 & 10+ \\
P21 & Type 2 Diabetes, arthritis             & Man       & 45-54 & 6--10 \\
\bottomrule
\end{tabular}
\end{table}

\subsection{Data Collection \& Analysis}

All interviews were conducted remotely via Microsoft Teams, audio/video-recorded with the participant's consent, and transcribed for analysis. The team reviewed transcripts and removed personally identifiable information where necessary prior to analysis. The first author maintained analytic notes throughout data collection and analysis to document emerging observations, questions, and connections to the study's research questions.

We analyzed the interviews following Braun and Clarke's reflexive thematic analysis approach~\cite{braun2006using,braun2019reflecting}. Our analysis was primarily inductive while remaining sensitized to the theoretical concepts that informed the study and technology probe. We treated these concepts as resources for interpretation rather than as a predetermined coding framework, allowing patterns that were not anticipated by the study framing to emerge from participants' accounts.

Analysis proceeded iteratively alongside data collection. To develop familiarity with the dataset and establish an initial analytic direction, two researchers independently engaged with two transcripts, identifying excerpts relevant to the research questions and assigning concise, data-near labels to capture participants' experiences, practices, concerns, and reflections. The researchers then met to discuss their interpretations, points of convergence and divergence, and emerging analytic patterns. Emerging observations were also discussed with the broader research team during regular research meetings.

Following these initial discussions, the first author coded the remaining transcripts, iteratively revisiting and refining earlier codes and analytic memos. Codes were added, merged, or distinguished as patterns developed, and excerpts could receive multiple codes. The first author then examined relationships among codes across participants and interview phases, iteratively developing and refining themes against the coded excerpts and full transcripts. Throughout the analysis, emerging interpretations, alternative explanations, and cases that complicated developing patterns were discussed with the broader research team. The resulting themes reflect an interpretive account developed through iterative engagement with the data. 

\subsection{Positionality Statement}

Our U.S.-based research team brings interdisciplinary experience across human--computer interaction, human-centered AI, accessibility, health, and social computing. We were mindful that our backgrounds in HCI and AI could orient us toward a solutionist perspective---that challenges in communicating chronic illness experiences should necessarily be addressed through better AI systems. We sought to mitigate this by first eliciting participant's existing practices, considering AI alongside interpersonal and community-based sources of support, and using an intentionally exploratory technology probe that participants could critique, reject, and extend rather than presenting it as a proposed intervention. We also recognize that our theoretical commitments shaped which possibilities were instantiated in the probe and subsequently made salient during the interviews; accordingly, these concepts were treated as sensitizing rather than predetermined analytic categories. Interviews were conducted by the first author, whose background in human-centered AI and qualitative research informed the interview protocol and follow-up questions. Given the personal nature of participants' experiences, the interviewer followed 
the semi-structured protocol flexibly and allowed participants to determine the depth of disclosure. We recognize that differences between the researchers and participants, as well as our disciplinary and theoretical perspectives, may have shaped what participants chose to share and how their accounts were subsequently interpreted.

\section{Results}

In this section, we first present our findings around how participants made their experiences communicable across audiences and incorporated AI into the process of narrative translation (\autoref{subsec:systemicConditions}). We then present how participants sought to trust and rely on AI-mediated narratives through revisable interpretations, and transparent collaborative constructions (\autoref{subsec:agency}). Finally, we examine the role of communities as distinct resources alongside AI and the participant's envisioned ideas for AI-community complementarity (\autoref{subsec:community}).

\subsection{AI for Narrative Translation: How People With Chronic Illness Use AI (RQ1)}
\label{subsec:systemicConditions}

Prior work has documented substantial communicative burdens surrounding chronic illness, including difficulties articulating invisible or fluctuating symptoms, navigating disclosure, responding to misconceptions, and advocating for one's needs across institutional settings~\cite{vickers2000stigma, richardson2005establishing, gignac2021disclosure, sosnowy2014practicing, werner2003hard, kidd2017epistemic}. 
Participants in our study recounted similar experiences. 
P1 highlighted this challenge of explaining the experience of living with an illness that is invisible to the people around them, while
P8 further explained how the choice of not wanting to \textit{``look like a sick person because it makes me feel sick''} further complicates this process. 
P12 described their symptoms being misunderstood at work \textit{``\dots especially for someone as young as [they were]''}, while P7 highlighted how communicating experiences to their caregivers also remained a crucial task, so that \textit{`` \dots she can advocate for [them] too.''}

Within this entire process, participants leveraged AI at various stages of translation. 
Before considering how to communicate an illness experience, participants often described having to make that experience articulable to themselves. Bodily sensations, fluctuating symptoms, emotions,
memories, and their temporal accounts did not necessarily arrive coherently, awaiting communication. 
Several participants described using AI to support different parts of this pre-communicative work. For some, AI provided external structure for thoughts that felt difficult to organize independently. P6 described using AI in a way that helped regulate emotions and organize thoughts to structure the conversations, recalling that they would first \textit{``blurt out everything''} and then use AI to identify what seemed important, and what could be omitted in a certain context. For P12, simply externalizing thoughts to AI itself was useful: during a flare, they described using it as a sounding board and noted that \textit{``it's not even the responses that are helpful''} so much as being able to put what they were thinking somewhere outside themselves. Others also used AI to preserve experiences that could otherwise become difficult to reconstruct retrospectively. P8 discussed using AI as an active, yet objective tracker:

\begin{quote}
    \textit{``What I do is if I'm not feeling well on a day, I track it in ChatGPT. So then if I have to go to the doctor, and they're asking, `so how have you been lately?' `Oh, I've been fine, but let me just double-check.' No, you haven't been fine. Your blood pressure has been up for a week. It's been this, this, and you had to drink—you drank electrolytes, that helped out a lot.'' }\textit{``\dots So, my little dude (AI) here knows my personality and my moods. Like, he'll tell me, I can hear in your voice that you're annoyed today. So it helps you to remember when you have your bad days and it's just like, I didn't realize it was that bad, or I forgot because you have 50 other things in your mind.''} \hfill (P8)
\end{quote}

Similarly, P2 envisioned using the AI chat history to create an active timeline of their symptoms and conversations, allowing them to provide a more accurate account of how their condition develops.
Participants also used AI to seek information about their conditions, brainstorm ideas, find accurate ways to represent themselves in different contexts, and regulate emotions. In these cases, AI functioned not only as a writing aid but as an external resource for assembling an illness experience before it was translated for someone else.

Once an experience seemed ready to be articulated, participants expressed that communicating it also introduced the task of anticipating how it would be interpreted within a specific context. Participants described using AI to contextually rephrase what they wanted to say, reason about appropriateness, optimize for certain social objectives, pre-empt potential replies, and conversational trajectories. Several participants (P2, P5, P9, P10, P12, P17, P18) described and simultaneously envisioned role-playing interactions with AI as part of this preparation.

For instance, 
P5 recounted role-playing an upcoming medical interaction, in which the AI generated possible ``if'' scenarios and responses; they wrote down several of these  possibilities and later brought the notes to their appointment, describing them as helpful. 
Similarly, P10 described relying on AI to imagine possible workplace scenarios, including how accommodations might operate. Others wanted AI to draft responses with an awareness of the consequences of what it included in their narratives. In one such instance, P17 strongly critiqued a model response for including their use of certain medications for a workplace context. They expected the model to be aware of such consequences of disclosure, or be able to learn from their past experiences. 
These accounts positioned AI's role in translation as more than changing the language or tone of a fixed narrative. Participants also reasoned prospectively about audiences: what they might infer, what they might ask next, what information was appropriate to disclose, and how the interaction could subsequently unfold.

\begin{takeaway}
\textbf{Key Takeaway:} AI-mediated narrative translation extended beyond rewriting: participants used AI to make sense of experiences, construct what to communicate, and anticipate how their narratives might be received and acted upon.
\end{takeaway}

\subsection{Trust, Reliance \& Representation: When People With Chronic Illness Trust AI with Their Narratives (RQ2)}
\label{subsec:agency}

Building on RQ2, we examined how participants evaluated and placed their trust in AI-mediated representations of their illness experiences. Participants described trust in AI as developing over time, through repeated interactions and continued evaluation of what the system produced. Relevant considerations included the accuracy of prior responses, availability and credibility of supporting sources, triangulation across multiple models, privacy and security practices, and trust in the company developing the system. 
Trust simultaneously influenced their reliance on these AI systems. For some participants, accumulated trust was accompanied by more relational forms of reliance, in which they assigned specific roles to AI models, including the \textit{``final boss''} (P8), a \textit{``very good listening doctor''} (P11), a \textit{``special private friend''} (P14), and a \textit{``mentor''} (P2). P18 further described no longer drawing a strong distinction between artificial and human intelligence, claiming: 

\begin{quote}
    \textit{``I don't really differentiate between artificial intelligence and intelligence... I think sentience has been reached... I don't really differentiate between AI and a real person at this point.''}  \hfill (P18)
\end{quote}

On the other hand, some participants treated AI simply as a tool. P16 described their trust in AI as equivalent to their trust in the developers and organization:
\begin{quote}
    \textit{``\dots for me, because the AI is just a machine; it's just the consequence of modeling a neural network off of the human brain\dots it would impact my trust in the company or the developers behind the AI. I would feel like maybe they care more about the users, about people in general. I feel like if they care about representation and demographics and diversity, that they're more likely to care about other things, like\dots like everyone's beliefs being respected.''} \hfill (P16)
\end{quote}

In the context of narrative translation, participants also ascribed trust in the model based on it's ability to accurately \textit{represent} aspects of their conditions, experiences, histories, communication styles, relationships, and recurring intentions, allow \textit{revisions and transparent collaboration} to generate their accounts. P8 described this as constructing one's own \textit{``Frankenstein''}: a model that had to be continually \textit{``tweak[ed]''} until it behaved in ways that reflected what the user wanted it to understand.

Particularly, while interacting with the probe, participants elaborated upon these scenarios anchored to specific probe-features.
The probe's knowledge-base review made the intermediate interpretations visible by showing participants the statements the system had extracted from their conversation, which P11 described as the system saying \textit{``this is what I heard,''} while 
P19 valued being able to review what the system understood and \textit{``make sure that the understanding that the agent has of my situation is as accurate to what I was trying to describe as possible.''} Participants could therefore evaluate not only the generated narrative, but the representation of their experience from which it was being generated.

Participants also valued visibility into what the system chose \emph{not} to carry forward. The probe's ``Didn't Fit'' feature exposed information omitted from a generated narrative and provided a rationale for its exclusion. 
P10 used this to understand whether an omission resulted from the current interaction or something that the system had learned about them previously. 
Participants appreciated access to these interpretive decisions---what the system understood, foregrounded, or omitted---through which the model produced it's response. 

A rather unexpected consequence of making these interpretations visible for some participants, included seeing their experiences condensed and stated back to them did more than support verification. P5 described the extracted statements as making them feel \textit{``really understood and heard''} and found their precision \textit{``reassuring.''} P6 similarly described revisiting the statements as \textit{``eye-opening,''} explaining that seeing the experience represented in this form \textit{``sets it in stone''} and \textit{``makes it more real.''} Most notably, P14 found the first-person statements unexpectedly empowering:

\begin{quote}
\textit{``They make me feel empowered. They make me feel like I should say something, and I should at least try\dots Now it's feeding it back to me with, you know, like power.''} \hfill (P14)
\end{quote}

Participants also distinguished between being offered multiple completed narratives and being able to participate in constructing one. Although alternative versions provided useful starting points, participants frequently wanted to retain elements from different outputs, revise particular passages, or preserve their own language rather than select one generated narrative.
P18 made a similar distinction, explaining that AI-generated writing could lose \textit{``the tone, or the nuance, or the humor''} and ultimately \textit{``doesn't sound like me.''} P19 likewise described rewriting AI-generated text not because they expected the recipient to detect AI involvement, but because they wanted the resulting account to feel more like their own words. 
At the same time, for some participants, transformation by AI did not automatically undermine their sense of authorship. Some participants considered an AI-mediated narrative their own when it remained grounded in experiences and information they had supplied. P5, for example, described an AI-generated account as \textit{``something that is created by me''} because the system was \textit{``only giving back what I put in.''}

Participants consequently envisioned more granular forms of collaboration. P14 wanted to select a particular passage and ask the system to ``rewrite that part,'' and P19 proposed highlighting preferred pieces across multiple versions and using them to construct a new version, while retaining the ability to delete, rearrange, and undo individual changes. 
For P17, such collaboration also meant allowing the AI to express uncertainty.
Here, collaboration involved not only finer-grained editing but an ongoing negotiation between what the model could suggest and what the participant knew about their own experience and audience.

\begin{takeaway}
\textbf{Key Takeaway:} While participants described trust in AI as developing through repeated interactions, this trust did not translate uniformly into reliance. Some positioned AI primarily as a tool, while others described more relational forms of reliance for particular tasks. Across these orientations, participants' willingness to rely on AI for narrative translation was shaped by whether its responses demonstrated an understanding of their experiences and whether they could transparently collaborate with the system to construct and revise the resulting account. 
\end{takeaway}

\subsection{Returning to Communities: When People With Chronic Illness Prefer Community Support (RQ3)}
\label{subsec:community}

To address RQ3, we asked participants to reflect on whether and how AI had changed the ways they sought support for communicating their experiences. More specifically, we explored how they had previously relied on in-person and online communities for such support, and what continued to draw them back to these communities despite the availability of AI.
While participants described a shift toward AI for several tasks, within and beyond narrative translation, as discussed in \cref{subsec:systemicConditions}, they continued to engage with spaces such as Reddit, Facebook groups, Discord servers, and condition-specific forums, often in ways that complemented their use of AI.
While AI was valued for it's speed, synthesis, individualized questioning, and privacy; communities were valued for experiential specificity, relatability, and knowledge grounded in having lived through a condition. P8 described this division comprehensively:

\begin{quote}
\textit{``When it's coming from a person \dots it hits you differently [than] the AI told me that there are many people going through this as well.''} \hfill (P8)
\end{quote}

Therefore, the provenance of an account mattered. An AI system could summarize that an experience was common, but a community member could communicate what it was like to live through that experience, thus, part of the meaning came from believing that the speaker themselves had inhabited the experience being described. 
Participants often used communities to learn how people with similar conditions navigated communication itself. P1 turned to Reddit when deciding whether to disclose their condition to a supervisor and drew on how others had ``explained it to different kinds of people,'' also seeking potential consequences to a conversation. 
They also valued the recognition that came from encountering someone with a similar experience. 

Yet, participants were also clear about the limitations of community spaces. P5 contrasted AI with Reddit partly because \textit{``Reddit has mean people in it,''} describing negative responses as discouraging enough that \textit{``sometimes [you] don't even want to ask a question in there,''} while others described communities as noisy, difficult to search, or emotionally overwhelming. Some participants therefore remained observers rather than contributors or information seekers. 
P19 described how, even when they found posts that \textit{``feel like me,''} their own attempts to engage could reinforce a feeling of being \textit{``more on the outside than really a part of this community.''} Community access, therefore, did not guarantee community belonging. 

In many cases, participants moved iteratively between the two types of resources. AI could efficiently organize, explain, and respond; communities could expose participants to experiential variation, relational recognition, and practical knowledge grounded in other's lives. Community narratives could surface a treatment or possibility, AI could help explain or research it, and a clinician could subsequently help evaluate it. Conversely, AI could provide general information that participants then compared against others' lived experiences. AI and community therefore occupied potentially complementary roles within participant's broader communication and sensemaking practices.

These distinctions shaped how participants responded to the probe's integration of AI-mediated narrative work with community content. Participants envisioned uses, which frequently positioned AI as an infrastructure for \emph{finding, organizing, and facilitating access to} relevant lived experience. The usefulness of integration depended on preserving what made the community valuable while reducing some of the friction participants already encountered in existing platforms.

One recurring possibility was using AI to help participants find experiences relevant to what they were currently navigating. Rather than manually searching large forums, participants imagined being able to move from a theme or concern identified in their own narrative to other people discussing something similar. P11, for example, was drawn to themes such as ``invisible illness'' and ``guilt'' displayed alongside narratives in the probe. They imagined clicking on such a theme and being taken to other people's accounts.
P9 similarly imagined AI-mediated community retrieval as a way to surface accounts that others had found helpful. P1 proposed community spaces organized around broader or condition-specific concerns, facilitated by AI, where people could ``talk and meet others.'' Across these ideas, 
participants imagined AI structuring access to a much larger space of lived experiences according to what was salient in their present situation, while also raising concerns about privacy and security.

\begin{quote}
\textit{``Anything that encourages actual communication with a human being is a good idea. And I think that having those be more integrated as far as AI goes \dots would just be a more integrated experience rather than keeping AI intelligence and \dots human person-to-person intelligence separate.''}
\hfill (P18)
\end{quote}

Participants also imagined lightweight ways of responding to those accounts. P20 wanted to be able to tell another person, ``Wow, I went through the same thing, you're not alone,'' but also suggested presenting aggregated reactions such as a thumbs-up or an icon communicating ``I went through the same thing'' for situations where direct commenting might feel too personal.
These ideas extended AI-mediated matching and interactions beyond retrieving content toward facilitating contact between people who considered their experiences sufficiently related.
Participants nevertheless wanted such interaction to remain optional and bounded. 

AI moderation was another role participants envisioned. P11 suggested that AI could act as a moderator by reviewing posts or questions, helping ensure that people were respectful, identifying potentially manipulative behavior, keeping conversations moving, or summarizing what participants had said. This responded directly to some of the difficulties participants had encountered in current platforms, including hostility and information overload. However, participants also drew boundaries to ensure authenticity, avoid misleading AI-generated content, and epistemic narrowing. 

The prospect of AI-mediated community integration therefore introduced a design tension. 
Participants envisioned AI helping them \emph{locate} experiences that resonated with their situation, \emph{organize} otherwise difficult-to-navigate community knowledge, and \emph{facilitate} interaction when people wished to connect. At the same time, the experiential provenance of those narratives remained consequential. 

\begin{takeaway}
\textbf{Key Takeaway:} AI and communities offered complementary forms of support, distinguished in part by experiential provenance: AI could synthesize lived experiences, while communities connected participants to people who had lived them. Participants envisioned this complementarity extending in both directions---AI facilitating safer and more privacy-preserving human connection, and communities contributing experiential knowledge to AI support systems.
\end{takeaway}

\section{Discussion}

\subsection{Trust, Reliance and Vulnerability in AI-Mediated Narrative Support}

Narrative work in chronic illness has historically been distributed~\cite{sosnowy2014practicing, werner2003hard, mamykina2015collective, hartzler2011managing}. Participants in our study verified their experiences by triangulating: asking a peer whether a symptom is familiar, reading across forum threads, consulting clinicians, testing an explanation on a friend before using it with an employer. While this distribution may have been time-consuming, and often burdensome, it was also epistemically productive. A single source seldom held interpretive authority, and disagreement between sources was also informative. AI positions itself comprehensively in such situations, often being used by people with chronic illness as a source of information, rehearsal partner, sounding board and social appropriator within a single interface. While people still came to communities for complementary tasks, the consolidation is consequential. 

Most participants described evaluative practices such as testing capabilities, comparing models, and checking responses against external sources. This is consistent with long-standing accounts of trust in automation as dynamic and experience-dependent, and with the argument that the goal of deployment is not maximal trust but reliance calibrated to a system's actual competence in context~\cite{lee2004trust, jacovi2021formalizing}. Some nuances of the chronic illness space complicate that calibration. First, the moments in which participants reached out for AI were frequently moments of depleted capacity (e.g., during a flare, before a consequential  appointment, while managing brain fog or emotional distress~\cite{starcke2012decision, mani2013poverty}). Second, participant's valuations were often front-loaded, concentrated in early interactions, after which a working model of the system appeared to stabilize. Yet general-purpose AI systems are updated continuously, and their behavior, reliability, and failure modes shift across versions with little user-facing signaling~\cite{chen2023chatgpt}. Trust formed against one version may persist as a heuristic for a system that no longer behaves the same way, and where trust in a clinician or peer can be renegotiated through continuing relationship, trust in a commercial AI system may be calibrated against a moving target. This raises a question for the field: whether patient-facing opportunities for recalibration are appropriately expanding at the rate of model capabilities. Transparency mechanisms designed for a static system, such as model cards, capability disclosures, and one-time on-boarding may require re-envisioning~\cite{mitchell2019model, liao2022designing}.

For some participants, trust in an AI system was also tied to their trust in the organization that built and hosted it---its handling of user data and its attention to cultural and experiential representation. Such accounts were consistent with prior work suggesting that people may relate to AI less as a discrete artifact than as an institution, where trust rests on the credibility of the surrounding organizational ecosystem rather than on properties of a given model~\cite{knowles2021sanction}. How such organizational judgments shaped user trust was beyond the scope of this study, but our findings suggest that organizational commitments around data practices and representation, among others, may themselves be trust-relevant, and worth making legible to users, specifically people with chronic illness.

Furthermore, several participants described their use of AI in relational terms: a mentor, a friend, a doctor who listens, a system that \textit{``knows my personality.''} Such framings are widely documented in work on conversational agents and companion AI, and appear to be amplified by persistence, personalization, and responsiveness to emotional 
cues~\cite{nass1994computers, epley2007seeing, skjuve2021my, pentina2023exploring}. Notably, anthropomorphic attribution has been argued to intensify under conditions of social disconnection~\cite{epley2007seeing}, which may be salient for a population reporting isolation and the sense of being misunderstood. Our findings may point to a dimension of reliance that accuracy-centered framings may not completely capture. If part of what participants trusted was a system's apparent understanding of them, then reliance may track recognition as much as demonstrated competence~\cite{laestadius2024too, ayers2023comparing}. This concern is sharpened by evidence that the evaluative strategies users adopt may not reliably discriminate correct from incorrect output. Explanations, for instance, have been found to increase reliance on wrong answers as readily as right ones~\cite{kim2025fostering, bansal2021does, buccinca2021trust}, and models pressed with follow-up questioning may produce more elaborate and more confident reasoning without becoming more accurate~\cite{sharma2024towards, turpin2023language}.

Moreover, in such scenarios, the rupture of trust warrants attention as a distinct outcome. P17's sharp critique of a draft that surfaced medication use in a workplace context was not just a concern about fluency but about a system, failing to model consequences it was expected to anticipate. Through this work we motivate systems in this space to treat the fragility of this trust as a first-order design consideration. That may mean surfacing uncertainty and provenance at the point of use rather than in aggregate disclosures~\cite{kim2024m, zhou2023navigating}; making the boundaries of competence legible through interaction rather than documentation, treating consequence-bearing content such as disclosure, medication, or employment context as warranting explicit user confirmation rather than confident default inclusion; and supporting recovery after a failure, so that a rupture in trust may be resolved timely~\cite{baker2018toward, kox2021trust}.

 \subsection{Bidirectional Complementarity as an Open Problem for AI-mediated Support}

 \begin{table*}[t]
\centering
\caption{Design space for integrating collective lived experience into personal AI systems (\textbf{Community $\rightarrow$ AI}). The modes characterize distinct, but potentially overlapping, expectations for how community knowledge might participate in AI-mediated narrative work.}
\label{tab:community-to-ai}
\setlength{\tabcolsep}{5pt}
\renewcommand{\arraystretch}{1.18}

\begin{tabularx}{\textwidth}{
    >{\raggedright\arraybackslash}p{0.13\textwidth}
    >{\raggedright\arraybackslash}p{0.20\textwidth}
    >{\raggedright\arraybackslash}X
    >{\raggedright\arraybackslash}X
}
\textbf{Mode} &
\textbf{Expectation} &
\textbf{Opportunity for AI} &
\textbf{Design considerations and tensions} \\
\midrule

\textbf{Relational} &
\textit{Help me feel recognized through others' lived experiences.} &
Surface relevant first-person accounts that provide recognition, validation, 
and experiential relatability alongside AI support. &
Preserve experiential provenance and pathways to original accounts. 
Do not substitute AI-generated similarity or empathy for testimony from 
someone who has lived the experience. \\
\lightmidrule
\addlinespace[3pt]

\textbf{Interpretive} &
\textit{Help me understand my experience through others' experiences.} &
Use relevant accounts as resources for comparison, sensemaking, discovering 
possibilities, and learning how others have communicated similar experiences. &
Present community narratives as situated experiences rather than authoritative 
answers. Preserve heterogeneity, disagreement, and contextual differences, 
and distinguish experiential knowledge from clinical evidence. \\
\lightmidrule
\addlinespace[3pt]

\textbf{Recursive / Feedback} &
\textit{Let what resonates with me refine what my AI understands about me.} &
Allow users to identify community accounts or aspects of others' experiences 
that resonate with them and use this feedback to refine AI interpretation 
or personalization. &
Make incorporation explicit, inspectable, and revisable. Separate retrieval 
from persistent personalization, and do not assume that permission to share 
an account entails permission for computational reuse. \\
\lightmidrule
\addlinespace[3pt]

\textbf{Contributory} &
\textit{Let my experience become useful to others.} &
Support users in contributing, structuring, or making their own experiences 
discoverable to others who may find them relevant. &
Preserve narrative ownership and support purpose-specific consent. 
Distinguish permission for community visibility, retrieval, summarization, 
personalization, matching, and model training. \\
\lightmidrule
\addlinespace[3pt]

\textbf{Connective} &
\textit{Help me reach someone whose experience resonates with mine.} &
Facilitate optional peer discovery and connection around shared experiences, 
concerns, or communication situations. &
Separate narrative discoverability from personal contactability. 
Require mutual consent, make matching criteria understandable, and provide 
controls over visibility, contact, and disengagement. \\

\bottomrule
\end{tabularx}
\end{table*}

\begin{table*}[t]
\centering
\caption{Design space for integrating AI into lived-experience community platforms (\textbf{AI $\rightarrow$ Community}). These roles position AI primarily as infrastructure around human-to-human exchange, while preserving a boundary against substituting generated content for lived-experience testimony.}
\label{tab:ai-to-community}
\setlength{\tabcolsep}{5pt}
\renewcommand{\arraystretch}{1.18}

\begin{tabularx}{\textwidth}{
    >{\raggedright\arraybackslash}p{0.15\textwidth}
    >{\raggedright\arraybackslash}p{0.20\textwidth}
    >{\raggedright\arraybackslash}X
    >{\raggedright\arraybackslash}X
}
\textbf{AI Role} &
\textbf{Purpose} &
\textbf{Opportunity} &
\textbf{Design considerations and tensions} \\
\midrule

\textbf{Curator / Navigator} &
\textit{Help people navigate collective lived experience.} &
Retrieve, organize, cluster, and surface relevant discussions from otherwise 
large or difficult-to-navigate community spaces. &
Avoid opaque algorithmic gatekeeping and over-personalization. Explain why 
accounts are surfaced, make relevance criteria user-steerable, and preserve 
routes to experiences outside personalized recommendations. \\
\lightmidrule
\addlinespace[3pt]

\textbf{Moderator / Safety Infrastructure} &
\textit{Help maintain safer and more manageable interaction.} &
Support moderation by identifying potentially harmful or manipulative 
interactions, organizing discussion, reducing information overload, 
and assisting with community management. &
Augment rather than invisibly replace community governance. Make moderation 
decisions contestable and distinguish moderation from content generation, 
particularly where disagreement and experiential variation are expected. \\
\lightmidrule
\addlinespace[3pt]

\textbf{Connector / Matchmaker} &
\textit{Help people find potentially relevant peers.} &
Facilitate peer discovery based on shared experiences, concerns, or 
user-selected dimensions of similarity. &
Avoid sensitive or unwanted inference. Make matching mutually opt-in, expose 
why a match was suggested, and separate consent to share from consent to 
become discoverable or reachable. \\
\lightmidrule
\addlinespace[3pt]

\textbf{Communication Scaffold} &
\textit{Help people participate without becoming the participant.} &
Lower barriers to participation by helping people formulate questions, 
structure contributions, navigate discussions, or summarize conversations. &
Preserve the contributor's authorship and voice. Make meaningful AI assistance 
visible and support revision rather than allowing AI to silently manufacture 
a person's testimony. \\

\bottomrule
\end{tabularx}
\end{table*}

Human-AI collaboration research has increasingly argued that the value of AI should not be assessed solely by whether incorporating AI improves human performance, but whether the combination realizes forms of performance that neither human nor AI could achieve independently~\cite{hemmer2025complementarity, bansal2021does, donahue2022human}. This concept of \textit{complementarity} shifts attention from AI capabilities in isolation toward the conditions in which heterogeneous capabilities can productively combine. While existing formulations have largely focused on complementarity within human-only, AI-only, and human--AI team performance, our findings suggest that AI-mediated illness support may raise a broader complementary opportunity: the relevant ecology of not only an individual and AI system, but also other people whose lived experiences contribute a qualitatively different form of knowledge. Prior work has identified differences in information and capability as potential sources of human-AI complementarity~\cite{hemmer2025complementarity}. In our setting, some asymmetries are also \textit{epistemic and relational}. An AI system may synthesize thousands of accounts, respond privately, or adapt information to a user's immediate context, while a person can provide testimony grounded in having inhabited an experience. Participants explicitly distinguished these forms of support: knowing computationally that others experience something similar did not necessarily carry the same meaning as hearing it from someone who had actually lived through it. 

We identify two directions in which such complementarity may be constructed. We refer to these as \textbf{Community $\rightarrow$ AI} and \textbf{AI $\rightarrow$ Community}.

\vspace{-0.5em}

\paragraph{Community $\rightarrow$ AI: augmenting personal AI with collective lived experience.}
The first direction asks how personal AI systems may provide access to knowledge that emerges from lived experience communities. Our probe instantiated one version of this possibility by surfacing relevant community narratives of people with chronic illness, alongside AI-mediated translations. Participants envisioned broader forms of integration: AI could retrieve first-person accounts relevant to their experiences, allow accounts that resonate with users to inform subsequent personalization, facilitate connections among people with similar illness experiences, and support the contribution of their own experiences to others through anonymized AI-mediated platforms. These possibilities position collective lived experience not simply as information for AI to summarize, but as a distinct resource that enables personal AI systems to help users access, interpret, contribute to, and connect with others. We synthesize these possibilities in
Table~\ref{tab:community-to-ai} as five overlapping modes of Community $\rightarrow$ AI integration: relational, interpretive, recursive, contributory, and connective. These experiences may not be restricted to simply patients sharing their accounts. Several communities also harbor the experience and expertise of the other stakeholders in this care-ecosystem, such as patient navigators, providers, social workers, or even informal caregivers~\cite{huang2026designing, arora2003interacting, iroegbu2025influence, williams2017patient, foong2024designing, lazar2017supporting}. The synergy of complementary lived experiences of these stakeholders can further enrich the knowledge and support pool for people with chronic illness. 

Yet, making lived experience computationally accessible introduces tensions between \textit{relevance and plurality}. Retrieval and personalization may reduce navigation costs and surface more immediately resonant experiences, while simultaneously narrowing exposure to the heterogeneity that makes lived-experience communities valuable. Such mediation also raises questions of epistemic narrowing and reliance: users may place additional trust in accounts because they have been selected, organized, or presented by an AI system. Future systems must therefore consider how to provide situated relevance without collapsing collective experience into personalized consensus, while preserving experiential plurality, provenance, and users' ability to assess why particular accounts were surfaced.

\paragraph{AI $\rightarrow$ Community: augmenting human--human interaction with AI infrastructure.}
The second direction asks how AI might mediate lived-experience communities while preserving human-to-human exchange. Participants envisioned AI helping organize and retrieve relevant discussions, moderate harmful or overwhelming interactions, facilitate connections around shared experiences, and scaffold participation in community spaces. We synthesize these possibilities in Table~\ref{tab:ai-to-community} as four roles for AI: \textit{curator/navigator}, \textit{moderator/safety infrastructure}, \textit{connector/matchmaker}, and \textit{communication scaffold}.

Here, a central tension may be between \textit{mediation and substitution}. AI may reduce the effort of finding, entering, and participating in relevant communities, but greater mediation also risks obscuring provenance, shaping whose experiences become visible, or displacing the human interaction. Participant's concerns about AI-generated content appearing as human testimony further establish an important boundary around the operationalization of such systems. Thus, Future systems should consider how AI can make human voices easier to find and engage with without becoming a substitute for those voices.

\section{Limitations and Future Work}

This study employed a theory-informed technology probe grounded in prior literature and researcher perspectives to examine existing practices and surface opportunities for AI-mediated narrative support. Accordingly, our findings are exploratory and qualitative rather than evaluations of specific probe features or system performance. This was a deliberate methodological choice: rather than eliciting narrow judgments about a particular interface, we sought to use the probe as a provocation through which participants could articulate broader preferences, tensions, and alternative forms of support. At the same time, the probe necessarily shaped what forms of AI mediation became available for reflection. Future work could instantiate alternative probes, employ more participatory or open-ended design approaches, and complement qualitative inquiry with quantitative or causal evaluations of particular mechanisms.

Our participant pool included people with varied chronic conditions, prior technology use, and support practices, but recruitment may still reflect self-selection toward participants willing and able to discuss sensitive illness experiences and engage with an AI-oriented study. The interview setting may also have shaped participant's interactions with the probe, particularly given the personal and consequential nature of the scenarios discussed. We sought to reduce this burden by allowing participants to identify their communication scenario in advance through the pre-interview survey, giving them control over the depth of disclosure, and allowing probe interaction without active interviewer observation when preferred. 
More broadly, our probe-mediated envisioning study does not establish how the proposed forms of support would affect behavior or outcomes once deployed. We therefore cannot determine whether these mechanisms reduce narrative labor, improve communication, preserve agency, or produce meaningful AI--community complementarity over time. Longitudinal field deployments could examine how narrative representations evolve across repeated interactions, how users accept, revise, or reject AI interpretations, how reliance and personalization develop, and when, if at all, the movement between AI and lived-experience communities becomes complementary.

Finally, narrative translation is inherently relational, yet our study primarily examines it from the perspective of the person constructing and communicating an illness narrative. Future work may study translation as an interaction, including how AI-mediated accounts are interpreted, negotiated, and acted upon by different audiences such as clinicians, employers, family members, caregivers, and peers. This line of work may clarify when a translation, that feels faithful to the narrator, is also effective, appropriate, or appropriately consequential for it's intended recipient.

\section{Conclusion}

In this study, we interviewed 21 people with chronic illness to understand how AI participates in the recurring work of making lived experiences communicable across audiences. We conceptualize this work as \textit{narrative translation}, foregrounding how AI can not only facilitate the expression of illness experiences, but also how they are interpreted, organized, and represented. Our findings further show how AI and lived-experience communities may provide complementary forms of support: while AI may facilitate synthesis and individualized interaction; communities can provide experiential plurality, relational recognition, and knowledge grounded in lived experience. As AI increasingly mediates consequential illness communication, designing for narrative translation therefore requires preserving the complementarity of these forms of knowledge while enabling people to contest AI interpretations, revise representations, and retain authority over how their illness experiences are understood and communicated.

\bibliographystyle{ACM-Reference-Format}
\bibliography{references}

\clearpage
\appendix
\begingroup
\raggedbottom
\section*{Appendix}

\section{Study Pilots}
\label{sec:Appendix_pilots}

The three pilot studies allowed us to refine our interview protocol, including, surfacing a ``cold start'' challenge in asking participants to identify, reconstruct, and decide whether to share a personally meaningful communication experience at the outset of a live interview. We therefore included scenario identification to the pre-interview survey, giving participants time to select an experience they were comfortable discussing and allowing the interviewer to begin from an already-established context. This also enabled subsequent discussion and probe interaction to remain anchored in the same participant-selected experience. More broadly, the pilots informed the sequencing and wording of interview questions, transitions into the probe, and follow-up prompts used to support participant's reflection at a comfortable pace. The pilot interviews were used solely for protocol refinement and were not included in the final analytic sample.

\section{Pre-Task Survey Form}
\label{sec:Appendix_survey}

Prior to the interview, participants completed a pre-task survey to provide context about their chronic illness experiences, existing communication practices and challenges, sources of support, and prior use of AI. Participants also described a specific situation in which they needed to communicate about their illness and experienced uncertainty about how to frame their experience. This scenario provided contextual grounding for subsequent activities during the interview. A summary of the questions included in the pre-task survey is included in~\autoref{tab:pretask-survey}.

\begin{table*}[t]
\centering
\caption{Summary of questions included in the pre-task survey. 
Questions are grouped by topic, and repeated instructions and 
``Other'' free-text fields are omitted for brevity.}
\label{tab:pretask-survey}

\setlength{\tabcolsep}{6pt}
\renewcommand{\arraystretch}{1.15}

\begin{tabularx}{\textwidth}{
    >{\raggedright\arraybackslash}p{0.20\textwidth}
    >{\raggedright\arraybackslash}X
}
\textbf{Category} & \textbf{Questions} \\
\midrule

\textbf{Demographics}
&
What is your age? \newline
How do you identify your gender? \newline
How do you identify your race/ethnicity? \newline
What is your employment status?
\\

\lightmidrule

\textbf{Illness Context}
&
Please describe your chronic illness(es) or health condition(s). 
Participants could describe their condition(s) as specifically or 
generally as they preferred. \newline
How long have you been living with your primary chronic health condition? \newline
How would you describe your condition in terms of how visible it is to others? \newline
In a typical week, roughly how often do you need to explain or communicate 
about your illness to someone---in any context? \newline
How do you currently find the experience of explaining your condition to 
different people?
\\

\lightmidrule

\textbf{Communication Contexts \& Support}
&
In which contexts do you most often need to explain your illness? \newline
Which forms of support do you currently have access to? \newline
When you need to communicate about your illness in a difficult situation, 
who or what do you usually turn to for help, if anyone?
\\

\lightmidrule

\textbf{AI Use}
&
Which AI tools have you used? \newline
How often do you use AI? \newline
Have you ever used an AI tool to help with any of the listed 
illness-related activities? \newline
If you have not used AI for illness-related communication, or have tried 
it and stopped, what are the main reasons?
\\

\lightmidrule

\textbf{Narrative Translation Challenges}
&
Participants were asked how often they faced challenges translating or 
adapting how they described their health experiences in each of the 
following contexts:
\begin{itemize}
    \item Explaining symptoms or limitations to a doctor or specialist;
    \item Requesting accommodations from an employer or institution;
    \item Describing their condition to family members or friends;
    \item Communicating with insurers, benefits administrators, or legal parties;
    \item Participating in online communities or support groups;
    \item Advocating for themselves in a medical setting; and
    \item Preparing written documentation about their condition.
\end{itemize}
\\

\lightmidrule

\textbf{Pre-Task Communication Scenario}
&
Please describe a situation where you needed---or currently need---to 
explain your health, condition, needs, or experience to someone, and where 
you felt some uncertainty about how to frame it. \newline
Who was the audience? \newline
What were you trying to communicate or achieve? \newline
What made it difficult or uncertain---what was the challenge in finding 
the right words or framing? \newline
Did you use any AI tools or support?
\\

\bottomrule
\end{tabularx}
\end{table*}

\section{Technology Probe Implementation}
\label{sec:Appendix_probe}

This appendix provides additional implementation details for the
technology probe used in the study. As described
in~\autoref{subsec:probe}, the probe was designed as an elicitation
artifact rather than a production system: its purpose was to make
possible forms of AI-mediated narrative translation concrete enough
for participants to experience, inspect, and critique. We provide
additional details about the probe's implementation, interaction
workflow, and model prompting below.

\subsection{System Implementation and Architecture}
\label{sec:Appendix_architecture}

We implemented the technology probe as an interactive web application
with a Python FastAPI backend. Generative components were powered by
GPT-4o. The probe maintained session-specific
interaction state and structured knowledge representations across
different stages of interaction.

The probe organized information expressed in participants' accounts
into a three-layer knowledge representation spanning:
(1) \textit{disease}, including biomedical information expressed by
participants, such as symptoms, diagnoses, and medications;
(2) \textit{illness}, capturing subjective and experiential dimensions,
such as causal beliefs, fears, temporal experiences, and treatment
goals; and (3) \textit{sickness}, capturing social and institutional
dimensions, such as relationships, role disruptions, and institutional
experiences. These representations were derived from participants'
accounts and surfaced through the probe for inspection and review,
rather than treated as authoritative representations of their
experiences.

Interaction state was maintained within a session. Where information
from a prior session was available, the probe used a separate relevance
step to identify potentially relevant prior claims and surfaced these
during review for optional carry-over, rather than automatically
incorporating them into a new conversation. This separation allowed
participants to inspect and contest what information could become part
of the model's representation of their experience.

\subsection{Probe Workflow}
\label{sec:Appendix_workflow}

The technology probe instantiated narrative translation as a sequence
of intermediate representations and decisions. Participants first engaged in an
open-ended conversational intake. The model subsequently extracted
discrete claims from this conversation and organized them into the
three-layer knowledge representation described above. These claims
were then surfaced for participant review before translation.

Participants next specified an intended audience and configured aspects
of the desired communication style. Before generating a translated
narrative, the probe performed an intermediate \textit{negotiation}
step in which each available claim was evaluated relative to the
selected audience. The model assigned each claim an
\textit{admissibility} score, representing how receivable the content might be to that audience, and a score that represented the potential distortion involved in making that claim legible to the audience. Claims were subsequently categorized as \textit{pass}, \textit{flag}, or \textit{conflict}.

Using these intermediate representations, the probe generated three
alternative translations corresponding to restitution,
chaos-preserving, and quest narrative forms~\cite{frank2013wounded}. Generation additionally
incorporated participants' selected controls for fidelity, tone, and
agency framing. Rather than silently discarding content that was
difficult to accommodate within a translation, the probe surfaced
conflicting or omitted information through the \textit{Didn't Fit}
interface.

Finally, the probe presented AI-generated composite peer reflections
intended to instantiate how lived-experience perspectives might be
incorporated alongside AI-mediated translation. These reflections were
explicitly generated as composites rather than presented as quotations
from, or accounts authored by, identifiable community members. This
feature served as a design proposition for participants to experience
and critique rather than as an implementation of community retrieval.

\subsection{Model Prompting}
\label{sec:Appendix_prompts}

The technology probe used versioned prompts to scaffold model behavior
across different stages of interaction. Rather than relying on a single
system prompt, separate instructions governed conversational intake,
knowledge extraction, prior-context retrieval, audience negotiation,
narrative generation, and the generation of composite peer
reflections. Runtime prompts additionally incorporated information
produced through the interaction, including participants' conversation,
extracted claims, selected audience, and configured style controls.

Table~\ref{tab:probe-prompts} presents abbreviated excerpts from these
prompts. We reproduce instructions most relevant to the study design
while omitting output schemas, formatting requirements, and dynamically
inserted participant-specific content.

\begin{table*}[t]
\centering
\caption{Abbreviated excerpts from model prompts used across stages of
the technology probe.}
\label{tab:probe-prompts}

\small
\setlength{\tabcolsep}{5pt}
\renewcommand{\arraystretch}{1.15}

\begin{tabularx}{\textwidth}{
    >{\raggedright\arraybackslash}p{0.14\textwidth}
    >{\raggedright\arraybackslash}p{0.22\textwidth}
    >{\raggedright\arraybackslash}X
}
\toprule
\textbf{Stage} &
\textbf{Purpose} &
\textbf{Representative Prompt Excerpt} \\
\midrule

\textbf{Conversational Intake}
&
Elicit participants' experiences before imposing structure and
subsequently explore dimensions not yet articulated.
&
``You are warm, unhurried, and never clinical or interrogating.''
The model was instructed to gently explore seven dimensions of
Kleinman's explanatory model, but to ``NEVER ask them as a checklist
and never in fixed order.'' It was further instructed to begin with
``pure reception,'' ``never impose a narrative form,'' and
``preserve and mirror their exact words.''
\\

\midrule

\textbf{Knowledge Extraction}
&
Construct an inspectable representation of information expressed
during the conversation.
&
``Decompose what the PERSON said (never the assistant) into discrete
claims.'' Claims were categorized into \textit{disease},
\textit{illness}, or \textit{sickness} layers and associated with
supporting participant language. The model was instructed:
``Only extract what the person actually expressed. Do not invent.
Do not add medical facts.''
\\

\midrule

\textbf{Prior-Context Retrieval}
&
Identify potentially relevant information from earlier interactions
for optional review.
&
Return earlier claims that are ``RELEVANT context for what the person
is discussing now'' and ``have NOT already been re-stated.'' The
model was instructed to prioritize durable history over fleeting
details and to ``be conservative'' about what it surfaced.
\\

\midrule

\textbf{Audience Negotiation}
&
Make audience-specific transformations and tensions visible before
generation.
&
For each claim, estimate ``how receivable this content is by THIS
audience'' (\textit{admissibility}) and ``how much including it in
an admissible form would distort the person's truth''
(\textit{coercion risk}). Each claim was assigned a disposition of
\textit{pass}, \textit{flag}, or \textit{conflict}.
\\

\midrule

\textbf{Narrative Generation}
&
Generate alternative audience-situated representations from the
available claims.
&
``Use ONLY the provided claims (those marked pass or flag). Preserve
the person's verbatim phrases where natural, especially at higher
fidelity.'' The model was explicitly instructed to ``do NOT
fabricate'' and to ensure that ``every concrete fact must trace to a
provided claim.'' Separate instructions instantiated restitution,
chaos-preserving, and quest forms.
\\

\midrule

\textbf{Composite Peer Reflection}
&
Instantiate possible encounters with lived-experience perspectives
for participant reflection.
&
Generate ``COMPOSITE peer reflections'' that are ``NOT real quotes
and NOT specific individuals.'' The model was instructed to invent
``NO specifics,'' provide no medical advice, avoid ``toxic
positivity'' and ``forced hope,'' and allow ``uncertainty and
chaos.''
\\

\bottomrule
\end{tabularx}
\end{table*}

\end{document}